\documentclass[twocolumn,showpacs,preprintnumbers,amsmath,amssymb]{revtex4}
\pdfoutput=1
\usepackage{graphicx}
\usepackage{color}

\newcommand{\sss}[1]{\scriptscriptstyle #1}

\begin{document}

\author{J. Ribassin}
\email{ribassin@apc.univ-paris7.fr}
\author{E. Huguet}
\email{huguet@apc.univ-paris7.fr} 
\author{K. Ganga}
\email{ganga@apc.univ-paris7.fr}
\affiliation{APC - Astroparticule et Cosmologie (UMR 7164),\\
Universit\'e Paris Diderot-Paris 7,\\
10, rue Alice Domon et L\'eonie Duquet F-75205 Paris Cedex 13, France}

\title{Simple fluid models for super-inflation in effective LQC and effects on the CMB B-modes}
\begin{abstract}
Loop quantum cosmology allows to replace the original singularity by a quantum bounce followed by a phase of fast expansion called super-inflation. In this paper, we use a simple analytic description of super-inflation using the equation of state of a fluid. We then derive the power spectrum of tensor perturbations produced. The effects on the B-mode polarization of the cosmological background radiation are discussed. For a large region of phase space, the spectrum should be distinguishable from the standard inflationary prediction for future observations. 
\end{abstract}

\maketitle

\section{Introduction}

The Loop Quantum Cosmology (LQC) \cite{BojowaldRevLQC08}, \cite{AshtekarSingh} (a recent review) is mainly the application to homogeneous spacetimes of the quantization methods of Loop Quantum Gravity (LQG) \cite{ThiemannLQGBook, RovelliLQGBook, AshtekarLewandowski}, a non-perturbative and background independent canonical quantization 
of general relativity. A central result of the theory is that the singular evolution encountered in the 
Wheeler-de~Witt theory no longer appears \cite{BojowaldLQCBounce, BojowaldLQCBounce2}. This is commonly interpreted as the replacement of the big bang singularity by a bounce. 
That result is a strong motivation to obtain quantitative predictions from LQC as a cosmological scenario. 
Since a complete quantum solution is not available at present, a constructive means to obtain some insights of the consequences of the phase close to the bounce is that of using effective
equations \cite{AshtekarSingh, AshtekarPawlowskiSingh2}. 
In such equations, the effects induced by the quantum nature of the geometry are taken into account
through corrective terms in classical equations.  In isotropic flat models (which are the only models 
considered here) two main corrections are used~: the "inverse volume correction" which comes from the
spectra of the quantum operator related to inverse triads, and the "holonomy correction" coming
from the quantization scheme in which holonomy of the connection are basic variables. 
The relative importance of the two type of corrections is still under discussion, it is then
usual to consider their effect separately. In the present work we focus on holonomy corrections and 
more specifically on the possible effects of the super-inflationary phase (in which the Hubble factor is
growing) \cite{BojowaldSInflationFirst, SinghSIGeneric} on the tensor modes of the 
power spectrum of the CMB. A number of study have already been devoted to tensor mode 
in LQC \cite{MulryneNunes, GrainBarrauCailleteauMielczarek}. Recently, effect of the big bounce
on the B-modes of the CMB has been considered \cite{GrainBarrauCailleteauMielczarek} 
(see also \cite{MaZhaoBrown} for a different approach), where the matter content of 
the universe was taken into account through a massive scalar field. 
In the present work we modelize the content of the universe during the phase of super-inflation using the simple equation of state 
$p = \omega \rho$, for which the modified Friedmann equation admits solutions in closed form for some relevant values of the constant $\omega$. We show that during the super-inflation 
that equation of state for a massless scalar field ($\omega =1$) is a very good
approximation to the massive scalar field in a fast roll regime ($\dot{\phi}^2 >> V(\phi)$).  
We compute in both cases (massless scalar and fast rolling massive scalar) the power 
spectrum after the phase of super-inflation including the holonomy corrections. We then estimate
the B-modes and discuss the sensitivity of the spectrum to the model considered.
 
In Sec. \ref{SecDynamics} we describe the dynamics the of a fluid governed by a simple equation of state
$p = \omega \rho$. We study how this equation compares with that of a massive scalar during the 
super-inflationary phase. The primordial and B-mode power spectra for the case $\omega =1$ are determined and discussed and compared to the standard inflationary prediction in Sec. \ref{SecPowerSpectra}. Except otherwise stated, the geometric units $\hbar = c = G = 1$ are used throughout the paper.

\section{Dynamics of the super-inflation from a simple fluid model}\label{SecDynamics}

We assume that the dynamics of the universe is described by the LQC modified 
Friedmann equation \cite{AshtekarPawlowskiSingh}:

\begin{equation}\label{EqFriedmannMod}
H^2 = \frac{\kappa}{3} \rho \left(1 - \frac{\rho}{\rho_c}\right),
\end{equation} 
where $\kappa = 8 \pi G$ and $\rho_c$ is the so-called critical density.  The modified Friedmann equation (\ref{EqFriedmannMod}) captures the ultraviolet corrections arising from the discrete quantum geometric effects. In particular, we see that when $\rho=\rho_c$ the theory departs drastically from general relativity: the Hubble parameter vanishes, and the Universe experiences a turnaround in the scale factor. The Big Bang singularity is therefore avoided and replaced by a big bounce.
The critical density can be expressed as a function of the so-called Barbero-Immirizi parameter $\gamma$ \cite{AshtekarSingh}:

\begin{equation}
\rho_c=\frac{\sqrt{3}}{16\pi^2G^2\gamma^3\hbar}\approx0.82\rho_{\text{Pl}}.
\end{equation}

Combining (\ref{EqFriedmannMod}) with the derivative of the continuity equation
$\dot{\rho} + H (\rho + p) =0$ leads to
\begin{equation*}
\dot{H} = - \frac{\kappa}{3}(\rho+p)\left(1 - 2\frac{\rho}{\rho_c}\right),
\end{equation*}
which shows that for $p\geqslant0$ the Hubble factor increases between the bounce ($\rho=\rho_c$) and 
$\rho=\rho_c/2$. This is the phase of super-inflation.

As for most non linear equations, it is often impossible to find an exact solution of the modified Friedmann equation.
However, with the generic equation of state $p = \omega \rho$, which, using the equation for the energy conservation, translates 
in $\rho = \xi a^{\sss -X}$, where $\xi$ and $X = 3(1+\omega)$ 
are constants and $a$ is the scale factor, the modified Friedmann equation becomes

\begin{equation}\label{EqDotaFromFriedmanAndState}
\dot{a}^2 = \frac{\kappa}{3} \xi a^{2-{\sss X}} \left(1 - \xi \frac{a^{\sss -X}}{\rho_c}\right).
\end{equation}
This equation can be solved literally for certain values of X,  in particular 
for $X=0$ (inflation or cosmological constant), $X=2$ (unknown fluid), $X=4$ (ultra-relativistic matter) and 
also (not so easily) for $X=6$ (massless scalar field).

The boundary condition for the modified Friedmann equation is given
at the bounce, where the density reaches the critical density. Thus 
$\xi = \rho_c a_0^X$, where $a_0$ is (chosen to be) the first eigenvalue of the scale factor which, according to \cite{init}, is:
\begin{equation*}
a_0 = \frac{3 \sqrt{\kappa}}{128 \sqrt{\pi} (\sqrt{2} - 1)}. 
\end{equation*}
Note that, we can also determine thanks to (\ref{EqDotaFromFriedmanAndState}) 
the value of the scale factor at the end of super-inflation ($\rho = \rho_c/2$) for $X >0$:
\begin{equation*}
a_{end} = 2^{\frac{1}{X}} a_0,
\end{equation*}
thus, the number of e-folds of super-inflation is simply given by:
\begin{equation*}
N = \frac{\ln(2)}{X} .
\end{equation*}
It is then clear that for $X>1$ super-inflation alone won't give a decent amount of efolds.

In order to compare with the the case of a single massive scalar field
one can rewrite the equation of state for the scalar field under the form 
$p=\omega\rho$ with $\omega$ and thus $X$ time dependant. Since the evolution of the scalar 
field is as usual governed by the Klein-Gordon equation
\begin{equation}\label{KG}
\ddot{\phi} + 3 H \dot{\phi} + m^2 \phi = 0.
\end{equation}

\begin{figure}[htbp]
\centering
\includegraphics[width=.4\textwidth]{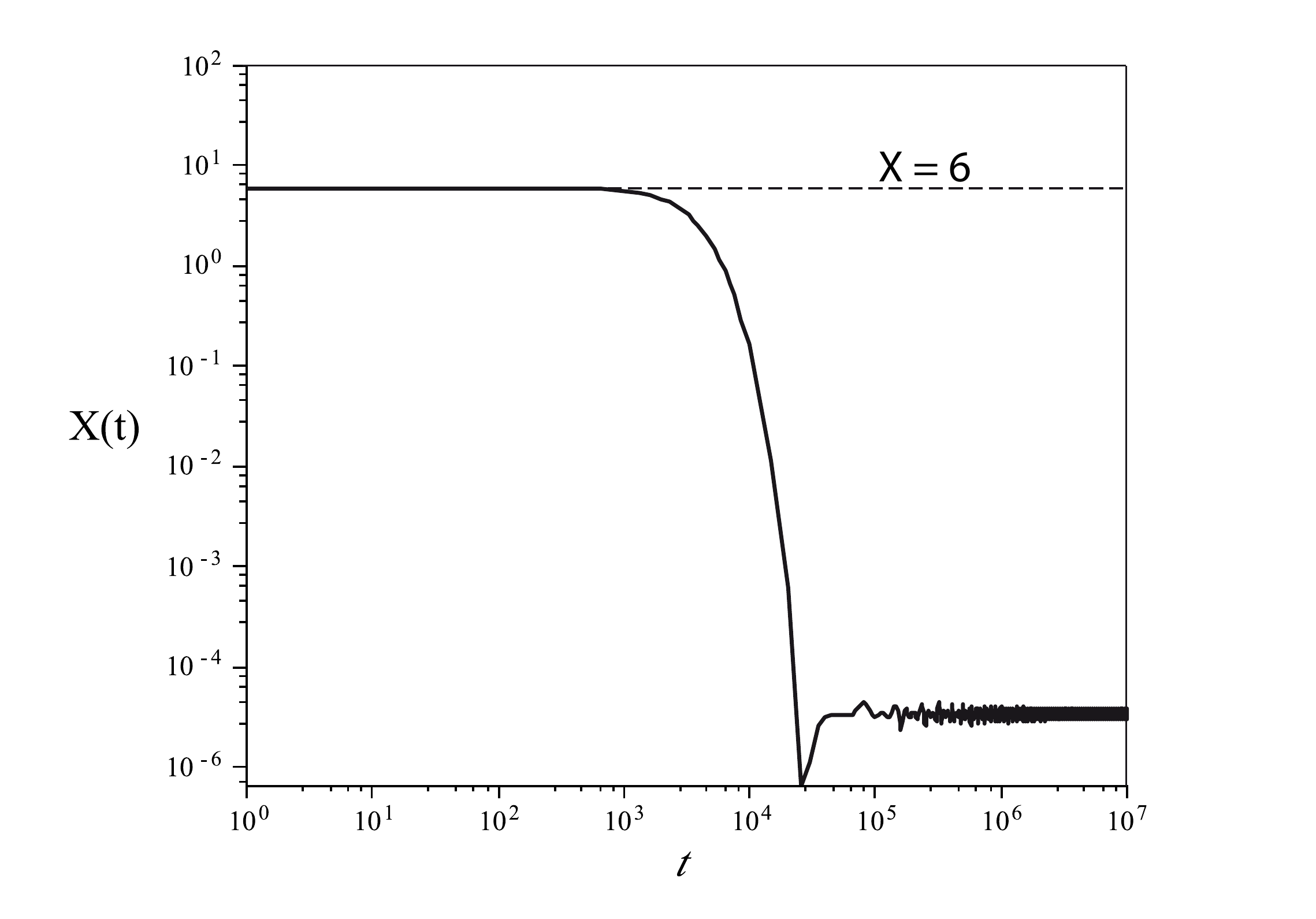}
\caption{Evolution of $X(t)$ for $m = 6.10^{-7}$ and $F_B = 10^{-9}$.}
\label{X(t)}
\end{figure}

By restricting the study to the case where the backreaction is negligible \cite{AshtekarSloan}, one can solve numerically this equation together with the modified Friedmann eq. (\ref{EqFriedmannMod}) to obtain 
the field $\phi(t)$. Then, from the expression of the pressure and the density of a scalar field, namely
\begin{align*}
p&=\frac{1}{2} \left(\dot{\phi}^2 - m^2\phi^2\right),\\
\rho&=\frac{1}{2} \left(\dot{\phi}^2 + m^2\phi^2\right),
\end{align*}
one obtains 
\begin{equation}\label{XdePhi}
X(t)  = \frac{6 \dot{\phi}^2}{\dot{\phi}^2 + m^2\phi^2}
\end{equation}
which is here a function of the time which depends on two physical parameters: $m$ and $F_B = m^2 \phi(0)^2/{2 \rho_c}$. It is pictured in Fig.~\ref{X(t)}.

\begin{figure}[htbp]
\centering
\includegraphics[width=.4\textwidth]{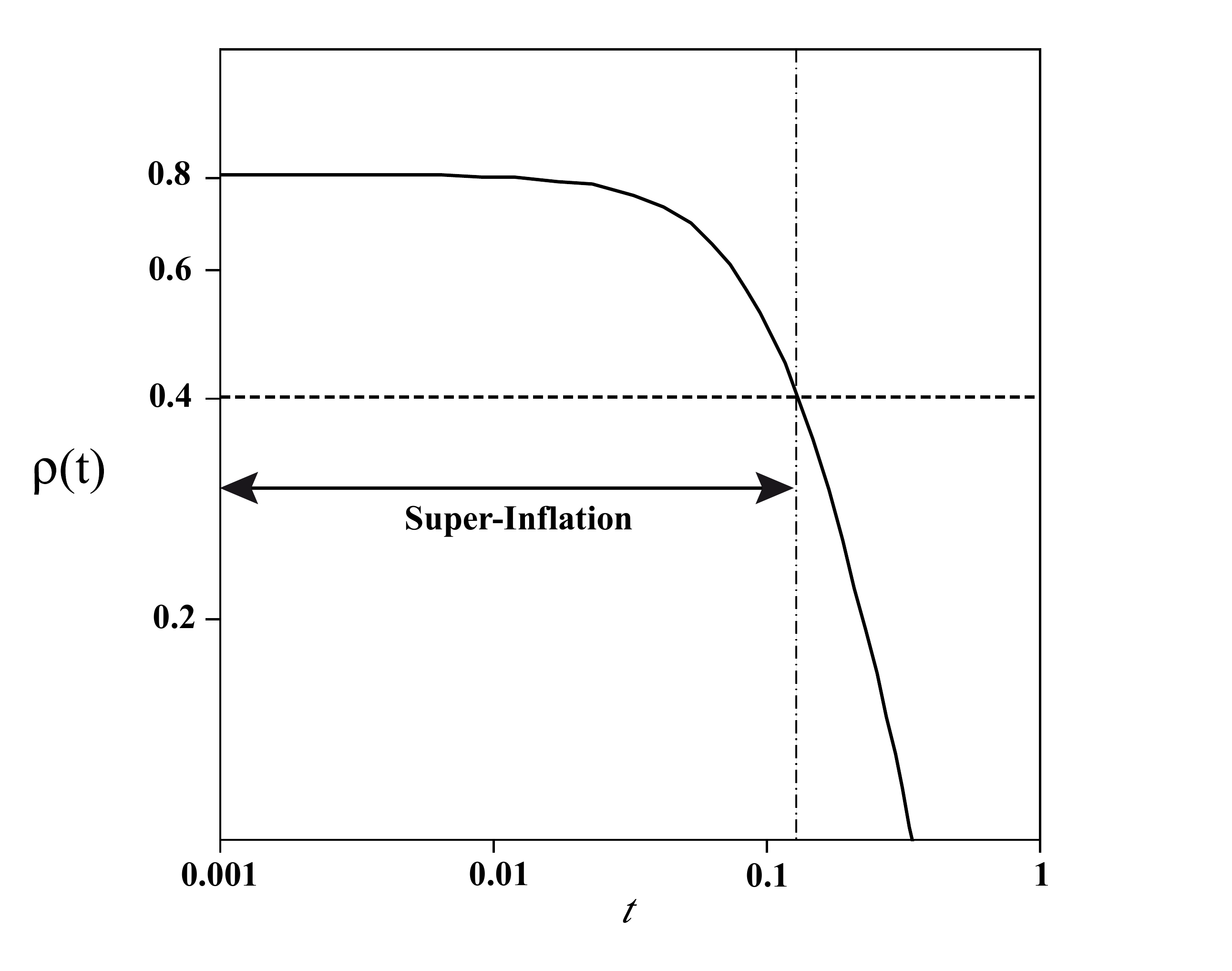}
\caption{Evolution of $\rho(t)$ for $m = 6.10^{-7}$ and $F_B = 10^{-9}$. The horizontal dashed line represents the value at which super-inflation ends. Here super-inflation lasts a bit more than 0.1 and thus $10^4$ times less than the phase with $X \approx 6$.}
\label{rho(t)}
\end{figure}

The evolution of $X(t)$ shows three main phases: a pre-inflationary phase ($0 < t < 10^3$) during which $X \approx 6$, an intermediate phase ($10^3  < t < 2.10^4$) during which $X(t)$ decreases and finally a phase of slow-roll inflation.
The length of the pre-inflationary phase is related to the value of $F_B$: if the potential energy at the bounce is small, it will take longer to reach a slow rolling regime.
During this pre-inflationary phase the dynamic of the massive scalar field can be approximated by the simple fluid  
$p= \omega \rho$ with $\omega = 1$ ($X = 6$). Such an equation of state corresponds to a massless scalar field. 
A more accurate description 
of this phase can be obtained with a fast rolling scalar field, i.e. such that $\dot{\phi}^2 >> V(\phi)$, 
with  equation of state $X(t) \approx 6$.
In what follows we will study these two approximations. 
Note that the time $t_{end}$ at which the super-inflation ends 
is easily determined from  $\rho(t)$ (Fig. \ref{rho(t)}) thanks to the relation $\rho(t_{\mathrm{end}}) = \rho_c/2$.
One observes that the super-inflation represents  a very small fraction of the pre-inflationary phase.

Let us now turn to the tensor modes. From the considerations of the previous paragraph, one can assume that the 
matter is described by a scalar and then consider the two approximations (massless and fast roll) for the pre-inflationary phase.
In order to obtain the power spectra of the perturbations, we have to solve the equation of propagation of the tensor modes, which can be written as
\begin{equation}
 u''_k(\eta) + \left(k^2 + m_{holo}^2(\eta) - \frac{a''(\eta)}{a(\eta)}\right) u_k(\eta) = 0,
\end{equation}
where $m_{holo}^2$ is the holonomy correction \cite{BojowaldHussain}
\begin{equation}
m_{holo}^2(\eta) = 2 \kappa a^2(\eta) \frac{\rho}{\rho_c} \left(\frac{2}{3} \rho - V(\phi)\right).
\end{equation}
Thanks to (\ref{XdePhi}), $V(\phi)$ can be written as function of X:
\begin{equation}
V = \xi a^{\sss -X} \left(1 - \frac{X}{6}\right),
\end{equation}
and thus
\begin{equation}
m_{holo}^2 = \frac{2 \kappa}{\rho_c} \xi^2 a^{2 - 2 {\sss X}} \frac{X - 2}{6}.
\end{equation}
We now determine the two quantities $m_{holo}^2$ and $a''(\eta)/a(\eta)$ for each approximation: free massless and 
fast rolling scalar. 

\subsection{Free Massless scalar field (X=6)}

The modified Friedmann equation can be written as:

\begin{equation}
\dot{(a^2)}^2 = \frac{4 \kappa \xi}{3 a^{2}}   ( 1 - \frac{\xi}{a^6 \rho_c}),
\end{equation}
its solution is then:

\begin{equation}
a^6(t) = \frac{\xi}{\rho_c} (1 +  \kappa \rho_c (t - d)^2),
\end{equation}
where $d$ is a constant determined by the initial conditions. At $t = 0$, one gets

\begin{equation}
a_0^6 = \frac{\xi}{\rho_c} = \frac{\xi}{\rho_c} (1 +  \kappa \rho_c d^2),
\end{equation}
which implies that $d = 0$. The conformal time is then given by

\begin{eqnarray}
\eta(t) & \equiv & \int \frac{d t}{a(t)}  \nonumber  \\ 
 & = & (\frac{\rho_c}{\xi})^\frac{1}{6} \;t \;_2F_1(\frac{1}{6}, \frac{1}{2}; \frac{3}{2}, - \kappa \rho_c t^2) \nonumber \\
 & = & \frac{1}{a_0} \;t + o(t^2),
\end{eqnarray}
where $_2F_1(n; d, z)$ is a hypergeometric function.

One can compute the terms of the propagation equation:

\begin{equation}
m_{holo}^2 = \frac{4 \kappa}{3 \rho_c} \xi^2 a^{-10} = \frac{4 \kappa}{3} \left(\frac{\rho_c^2}{\xi}\right)^\frac{1}{3}   \left(1 +  \kappa \rho_c t^2\right)^{- \frac{5}{3}}
\end{equation}
and 

\begin{equation}
\frac{a''}{a} = - \frac{\kappa \xi^{1/3}}{27}\; \rho_c^{2/3} \;(2 \kappa \rho_c t^2 - 3)\; \frac{3 \kappa \rho_c t^2 + \kappa \rho_c t + 3}{ (1 + \kappa \rho_c t^2)^\frac{8}{3}}.
\end{equation}

\subsection{Fast rolling scalar field ($X \approx 6$)}

For a fast rolling scalar field ($\frac{1}{2} \dot{\phi}^2 >> V(\phi)$), the Friedman (\ref{EqFriedmannMod}) and Klein Gordon (\ref{KG}) equations reduce to:

\begin{equation}
H^2 = \frac{\kappa}{6} \dot{\phi}^2 (1 - \frac{\dot{\phi}}{2 \rho_c}),
\end{equation}
\begin{equation}
\ddot{\phi} + 3 H \dot{\phi} = 0.
\end{equation}
Thus
\begin{equation}
\ddot{\phi}^2 = \frac{\kappa}{6} \dot{\phi}^4 (1 - \frac{\dot{\phi}}{2 \rho_c}),
\end{equation}
whose solution is

\begin{equation}
\dot{\phi}(t) = \sqrt{\frac{6 \rho_c}{3 + \kappa \rho _c (t - A)^2}}
\end{equation}
where $A$ is a constant fixed by the initial conditions. At $t = 0$, $\rho = \rho_c = \frac{1}{2} \dot{\phi}^2$ and so one gets

\begin{equation}
\dot{\phi}(0) = \sqrt{2 \rho_c} = \sqrt{\frac{6 \rho_c}{3 + \kappa \rho _c  A^2}}
\end{equation}
which implies that $A = 0$. Using the Friedman equation, one can then derive the Hubble parameter:

\begin{equation}
H(t) = -\frac{1}{6} \frac{\kappa \rho _c t}{3 + \kappa \rho _c t^2}
\end{equation}
and since $H = {\dot{a}}/{a}$, the scale parameter is:

\begin{equation}
a(t) = a_0 (\frac{1}{1 + \frac{\kappa \rho_c}{3}  t^2})^\frac{1}{6}.
\end{equation}

As in the previous case, the conformal time takes the form

\begin{eqnarray}
\eta(t) & \equiv & \int \frac{d t}{a(t)}  \nonumber  \\ 
 & = &  \frac{1}{a_0} \;t \;_2F_1(\frac{1}{6}, \frac{1}{2}; \frac{3}{2}, - \kappa \rho_c t^2) \nonumber \\
 & = & \frac{1}{a_0} \;t + o(t^2).
\end{eqnarray}
We can compute the terms of the propagation equation:

\begin{equation}
m_{holo}^2 = \frac{\kappa}{3 \rho_c} a^{2} \dot{\phi}^4 = 3^\frac{1}{3} \;\frac{12 \; \kappa \rho_c a_0^2}{(3 + \kappa \rho_c t^2)^\frac{7}{3}}
\end{equation}
and 

\begin{equation}
\frac{a''}{a} = 3^\frac{1}{3} \; \frac{\kappa \rho_c a_0^2}{9} \; \frac{5 \kappa \rho_c t^2 - 9}{(3 + \kappa \rho_c t^2)^\frac{7}{3}},
\end{equation}
which are, contrary to the conformal time, slighty different from those obtained for the massless scalar field.

\section{Power spectra}\label{SecPowerSpectra}

\subsection{super-inflation and beyond}

Let us consider the power spectra of the super-inflation alone. For both our modelisations, the equation of propagation of the tensor modes admits no exact solution and was thus solved numerically using a Runge-Kutta method. the power spectra could then be expressed as:

\begin{equation}
P(k) = \frac{k^3}{2 \pi^2} \left \lvert \frac{u_k(\eta_{end})}{a(\eta_{end})} \right \rvert^2,
\end{equation}
where $\eta_{end}$ corresponds to the end of the pre-inflationary phase ($t_{end} \approx 10^3$, see Fig. \ref{X(t)} ). Thus, we are not only computing the pertubations produced during super-inflation but also those emerging from the intermediate phase that takes place just before the slow-roll regime.

\begin{figure}[htbp]
\centering
\includegraphics[width=.5\textwidth]{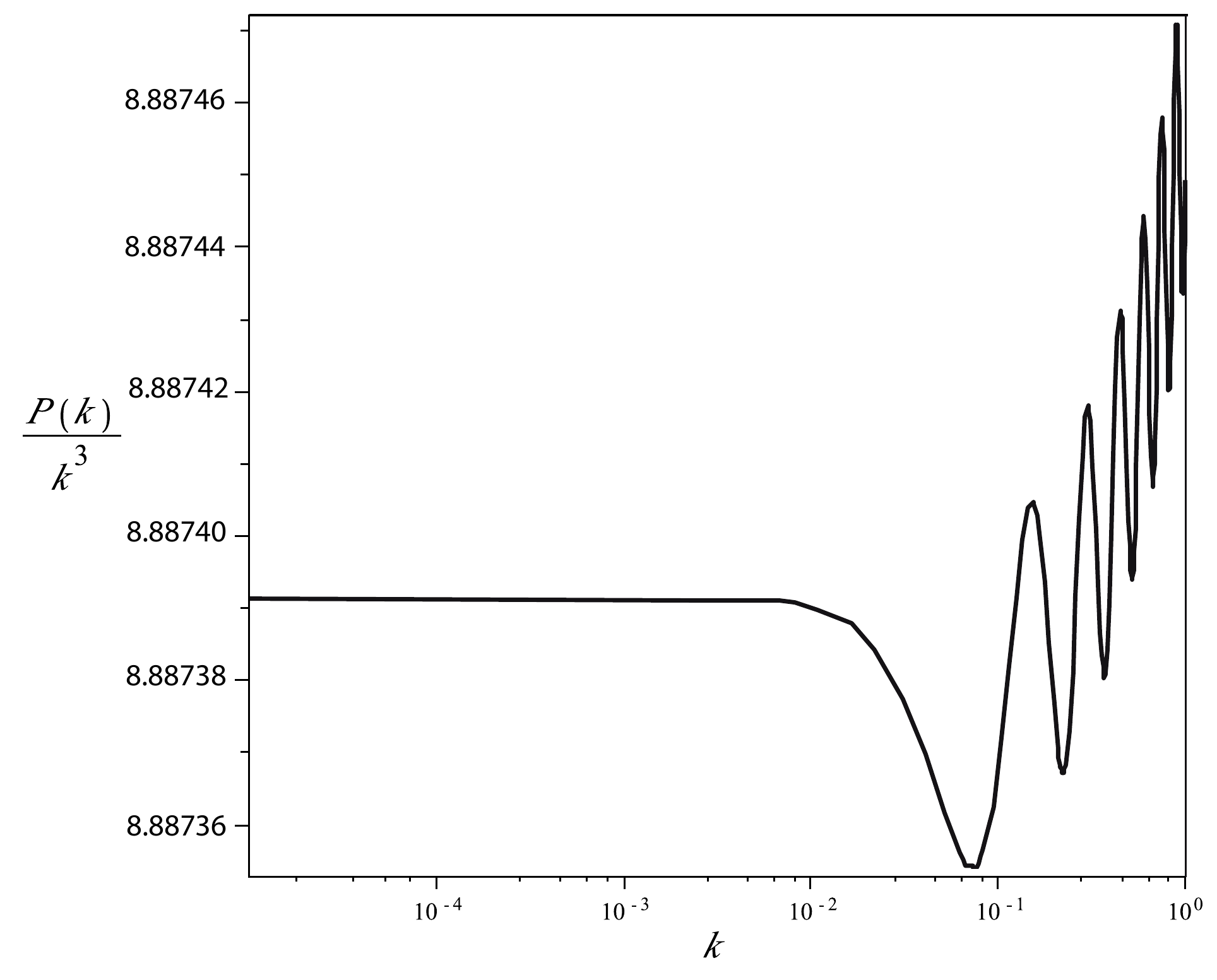}
\caption{Power spectrum of the pre-inflationary phase modelized as a massless scalar field ($X=6$).}
\label{X=6}
\end{figure}

\begin{figure}[htbp]
\centering
\includegraphics[width=.5\textwidth]{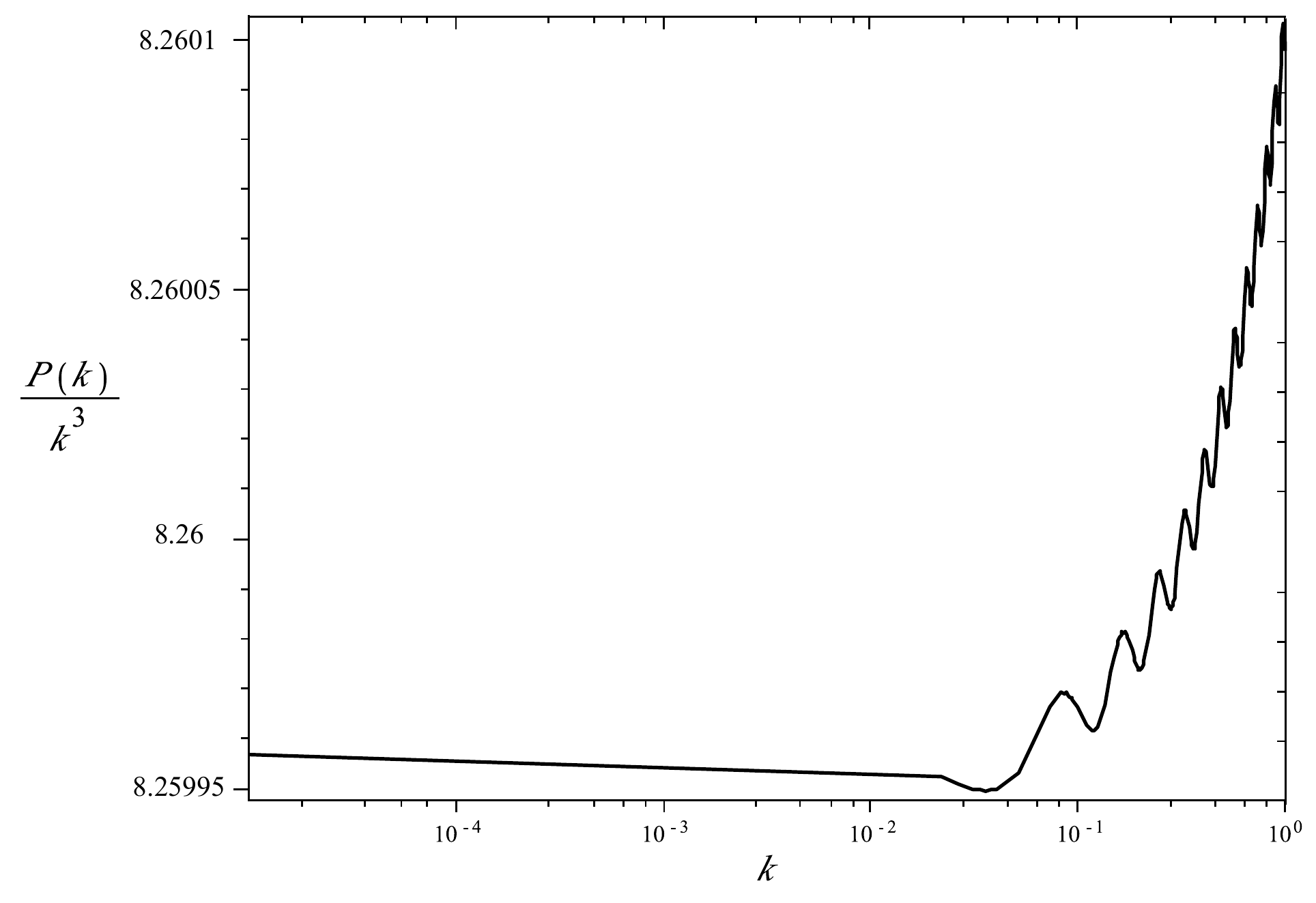}
\caption{Power spectrum of the pre-inflationary phase modelized as a fast rolling scalar field ($X \approx 6$).}
\label{fastroll}
\end{figure}

\begin{figure}[htbp]
\centering
\includegraphics[width=.55\textwidth]{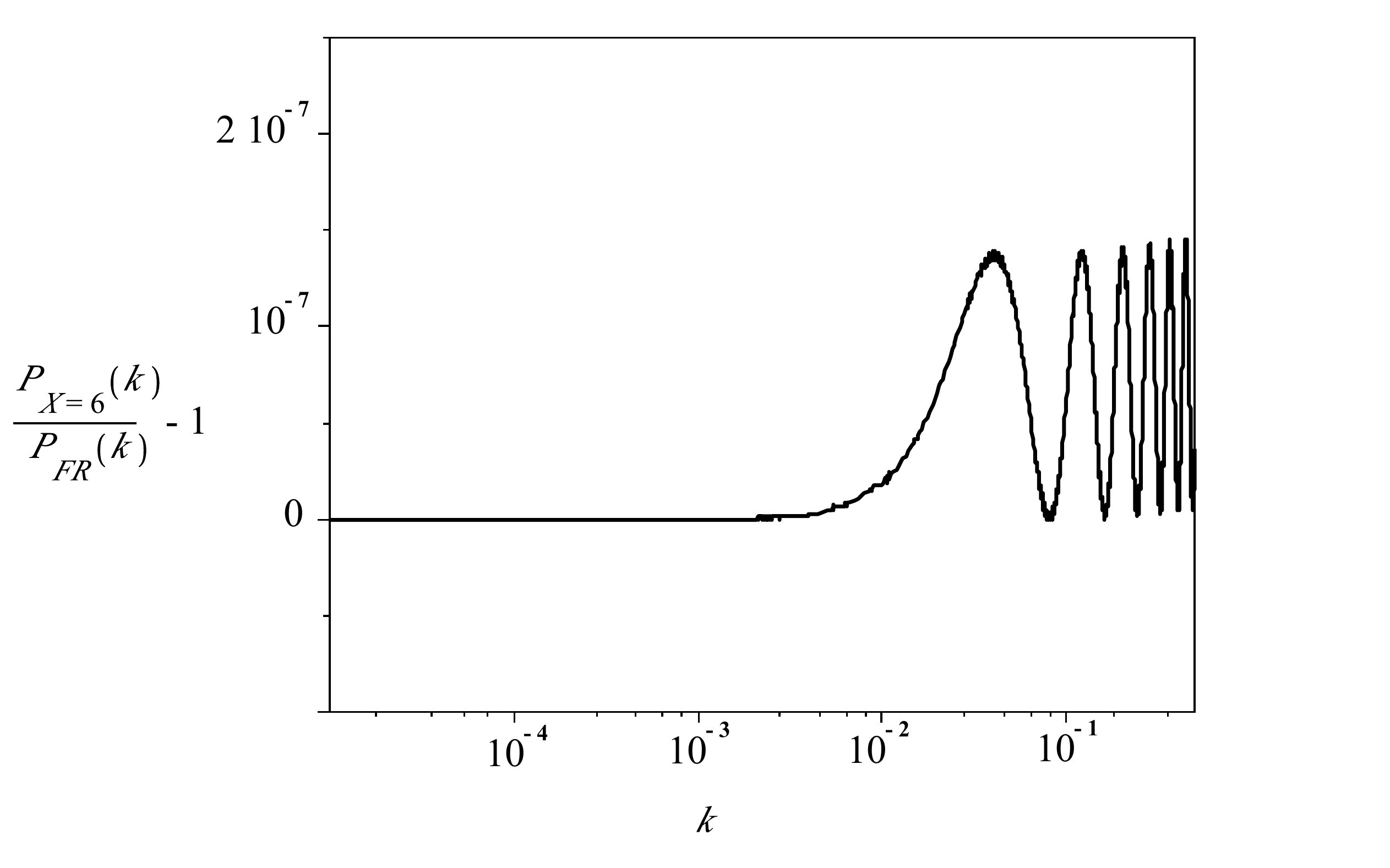}
\caption{Comparison between the power spectra in the fast-roll and $X=6$ cases.}
\label{X=6_FS}
\end{figure}

In both cases (Fig. \ref{X=6} and \ref{fastroll}), the slope at large scales grows like $k^3$ and the spectra begin to oscillate very slightly around $k = 10^{-2}$. The minor differences between the two spectra can be ignored (see Fig. \ref{X=6_FS}) as they would be undistinguishable at the scale of the CMB. We therefore conclude that a simple massless scalar field is suitable to describe the perturbations produced by super-inflation. 

\subsection{Complete spectrum of perturbations}

\begin{figure}[htbp]
\centering
\includegraphics[width=.45\textwidth]{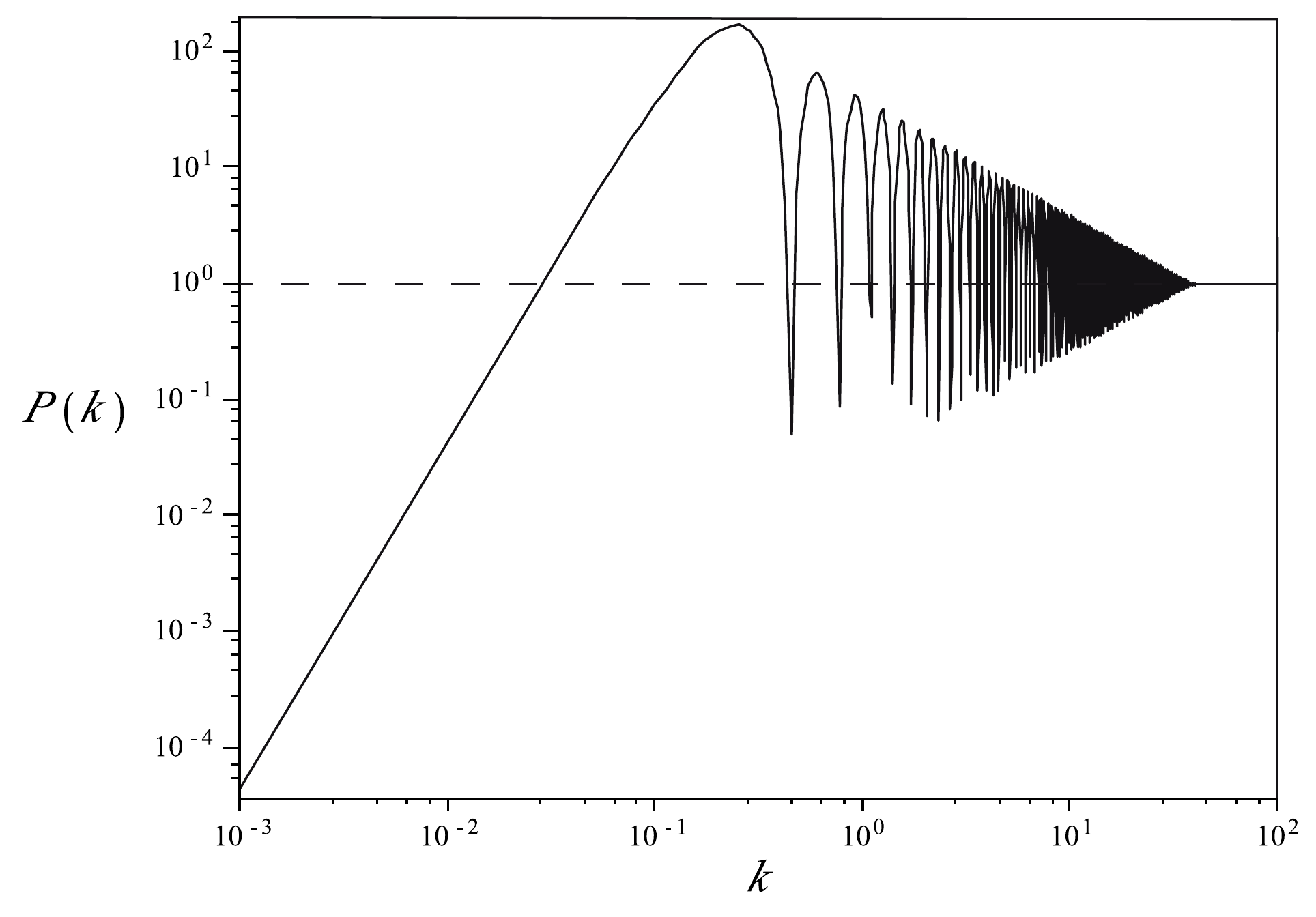}
\caption{Power spectrum of the pre-inflationary phase followed by inflation for $m = 6.10^{-7}$ and $F_B = 10^{-9}$. As usual only the super-Hubble modes are considered.}
\label{big9}
\end{figure}

\begin{figure}[htbp]
\centering
\includegraphics[width=.55\textwidth]{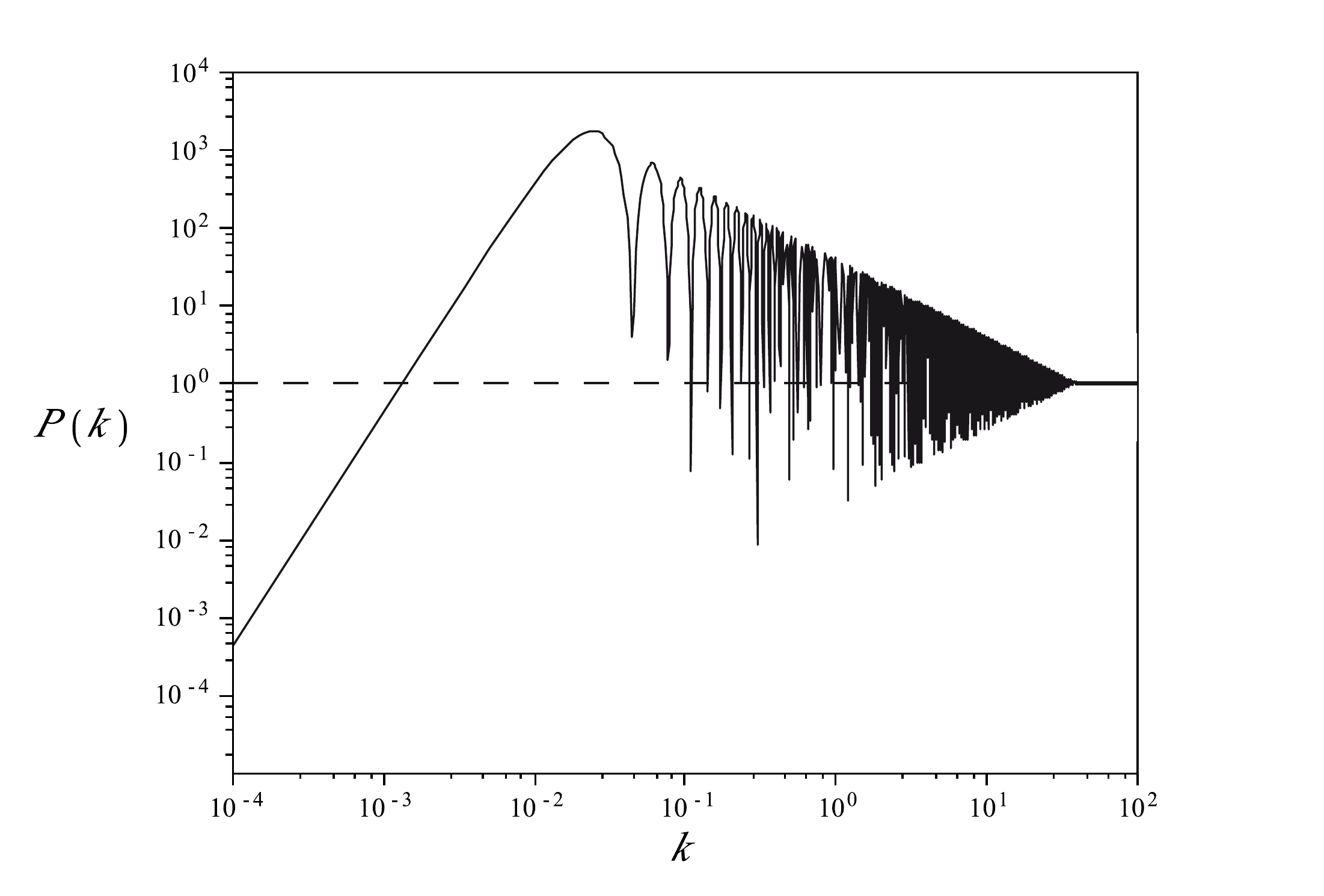}
\caption{Power spectrum of the pre-inflationary phase followed by inflation for $m = 6.10^{-7}$ and $F_B = 10^{-11}$. As usual only the super-Hubble modes are considered.}
\label{big11}
\end{figure}

To obtain the power spectrum corresponding to the total primordial epoch we have to connect the perturbations that arise from the pre-inflationary phase to a phase of standard slow-roll inflation. To this end we assume that we neglect the perturbations produced during the transition between the two main phases. We also consider that the quantum corrections are negligible during inflation since $\rho << \rho_c$ \cite{AshtekarSloan}. The perturbations produced during inflation thus take the usual form

\begin{equation}
v_k(\eta) = - C_1 \sqrt{\frac{\pi \eta}{4}} \; H_{\frac{3}{2}}^{(1)}(- k \eta)  - C_2 \sqrt{\frac{\pi \eta}{4}} \; H_{\frac{3}{2}}^{(2)}(- k \eta),
\end{equation}
where $C_1$ and $C_2$ are constants and $H_{\frac{3}{2}}^{(1)}$ and $H_{\frac{3}{2}}^{(2)}$ Hankel functions of the first and second kind. At the end of the first phase, we impose the continuity of the perturbations and of their first derivative:

\begin{eqnarray}
u_k(\eta_{end}) &=& v_k(\eta_{end}), \nonumber \\
u_k'(\eta_{end}) &=& v_k'(\eta_{end}).
\end{eqnarray}
We can then produce the power spectrum of $v_k$ at the "end" of inflation (so that the whole phase would yield 70 e-foldings). As in the case of pre-inflationary phase alone, the power spectra (Fig. \ref{big9} and \ref{big11}) still behave like $k^3$ at large scales. For large k the spectra are nearly scale invariant as one would expect with inflation. Some oscillations appear between those two regimes. Lowering the value of the parameter $F_B$ increases  the amplitude of these oscillations which thus appear as the consequence of the bounce. As seen in Sec. \ref{SecDynamics}, decreasing $F_B$ increases the length of the pre-inflationary phase, thus giving more time to the holonomy correction to modify the spectrum of perturbations.

\subsection{B-mode angular power spectrum}

Since we have computed the power spectrum of the tensor perturbations, we are now able to produce the primordial part of the B-mode 
angular power spectrum of polarization of the CMB. We do not know the modification of the spectrum of scalar perturbations produced 
by the holonomy correction, we will thus assume that scalar perturbations are given by the usual nearly scale invariant spectrum. 
This will allow us to produce the part of the spectrum that emerges from the conversion from E-modes to B-modes through lensing. 
Moreover, the holonomy correction is an ultraviolet correction while the lensed part of the BB angular spectrum is supposed to 
dominate mostly at small angular scales.

\begin{figure}[htbp]
\centering
\includegraphics[width=.5\textwidth]{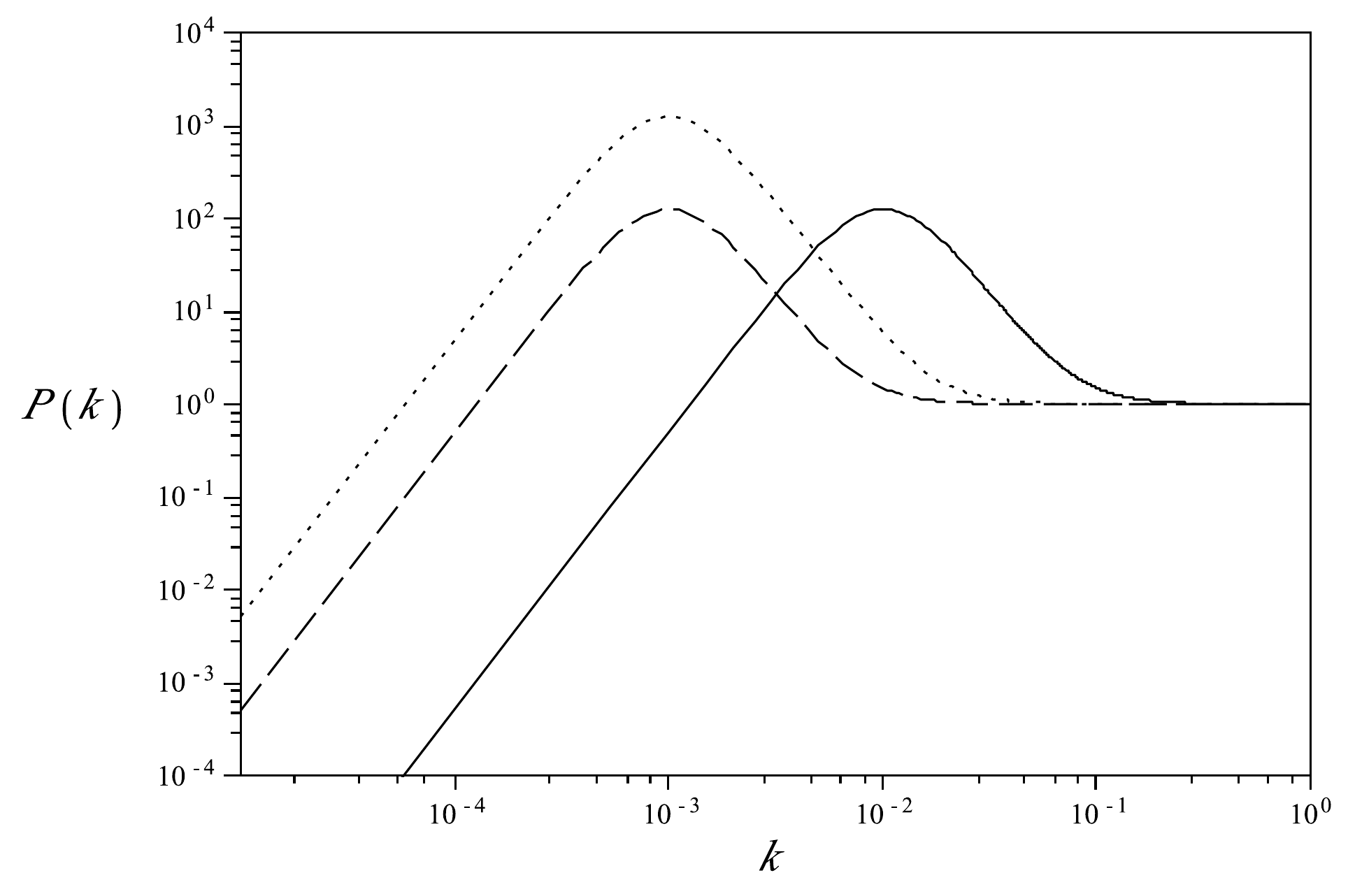}
\caption{Parametrization of the primordial power spectrum given by equation (\ref{parametrization}) 
with $(k_0, R) = (10^{-2}, 500)$ (solid), $(k_0, R) = (10^{-3}, 500)$ (dashed) and $(k_0, R) = (10^{-3}, 5000)$ (dotted).}
\label{param}
\end{figure}

The B-mode power spectra are obtained using CAMB \cite{CAMB}. For the cosmological parameters, we took those 
given for $\Lambda CDM$ by WMAP+BAO+H0 \cite{WMAP7}. As input for the primordial power spectrum we used a simple parametrization 
of  the previous spectrum inspired by \cite{GrainBarrauCailleteauMielczarek} 

\begin{equation}
\label{parametrization}
P(k) = \left(\frac{H}{2 \pi}\right)^2 \; \frac{1}{1 + (\frac{k_0}{k})^3} \;\left(1 + \frac{R}{1 + (\frac{k}{k_0})^3}\right).
\end{equation}
The parameters $k_0$ and $R$ respectively determine the position and amplitude of the transition between 
two regimes (Fig. \ref{param}): $P(k) \propto k^3$ at large scale and the spectrum is scale invariant at lower scales. 
The transition between those two regimes presents a bump. We will not take into account the oscillations of the 
primordial spectrum in the rest of the computation and only focus on the envelope. The parameter $R$ is closely 
related to the parameter $F_B$: high values of $R$ correspond to small values of the potential energy of the scalar field at the bounce. 
The link between the phenomenological parameter $k_0$ and the physical parameters is less simple since the position of the bump depends
 on both $F_B$ and the length of the slow-roll phase.

%

%


%

\begin{figure}[htbp]
\centering
\includegraphics[width=0.5\textwidth]{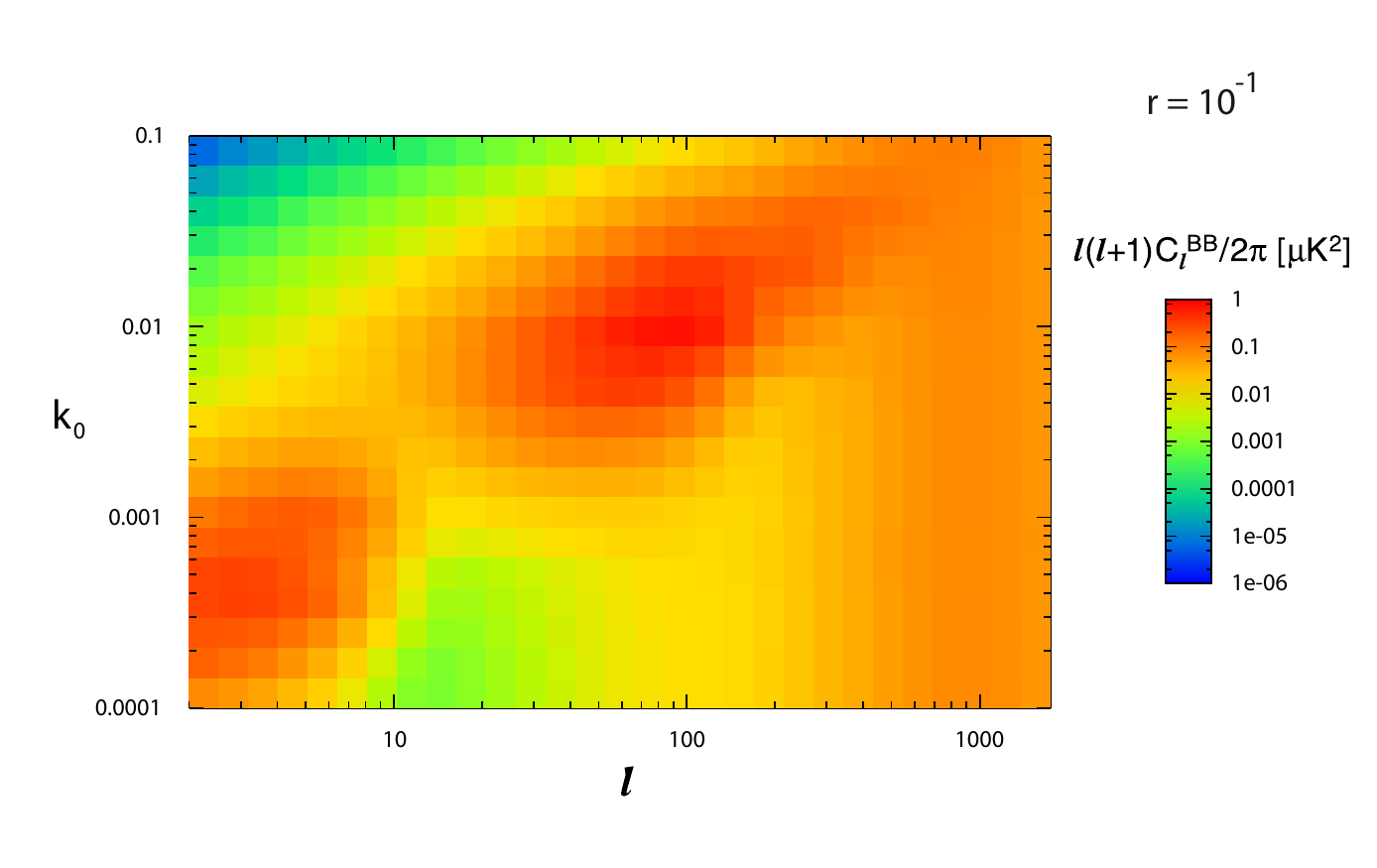}
\caption{Angular power spectrum of the B modes as a function of $k_0$ with a tensor-scalar ratio of $10^{-1}$ and $R=500$.}
\label{BB-1}
\end{figure}

\begin{figure}[htbp]
\centering
\includegraphics[width=0.5\textwidth]{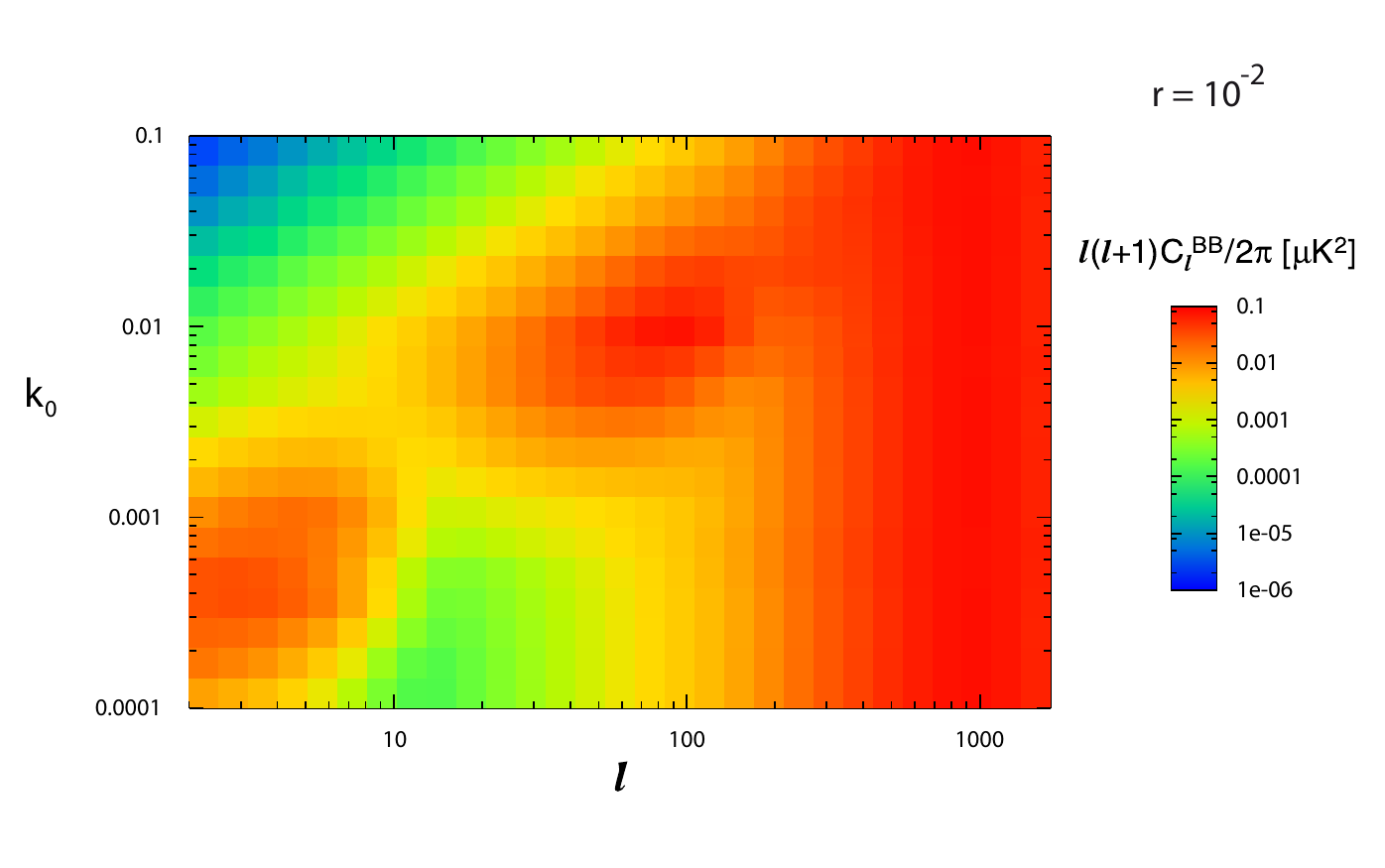}
\caption{Angular power spectrum of the B modes as a function of $k_0$ with a tensor-scalar ratio of $10^{-2}$ and $R=500$.}
\label{BB-2}
\end{figure}

The resulting B-mode power spectra are presented as a function of the parameter $k_0$ in Fig. \ref{BB-1} and \ref{BB-2} 
with tensor-scalar ratios of $10^{-1}$ and $10^{-2}$ respectively. At large multipole, the spectra are dominated by the 
contribution of the lensing as in the usual scenario and are thus independant of $k_0$. However, the spectra also reach a 
maximum in two regions: at large angular scales ($\ell < 10$) for $k_0 < 10^{-3}$ and at intermediate scales ($20 < \ell < 200$) for 
values of $k_0$ around $10^{-2}$. The latter case is the most interesting from an observational point of view since it coincides with 
the beginning of the observationnal window of ground based experiments like BICEP \cite{bicep}.

\begin{figure}[htbp]
\centering
\includegraphics[width=0.5\textwidth]{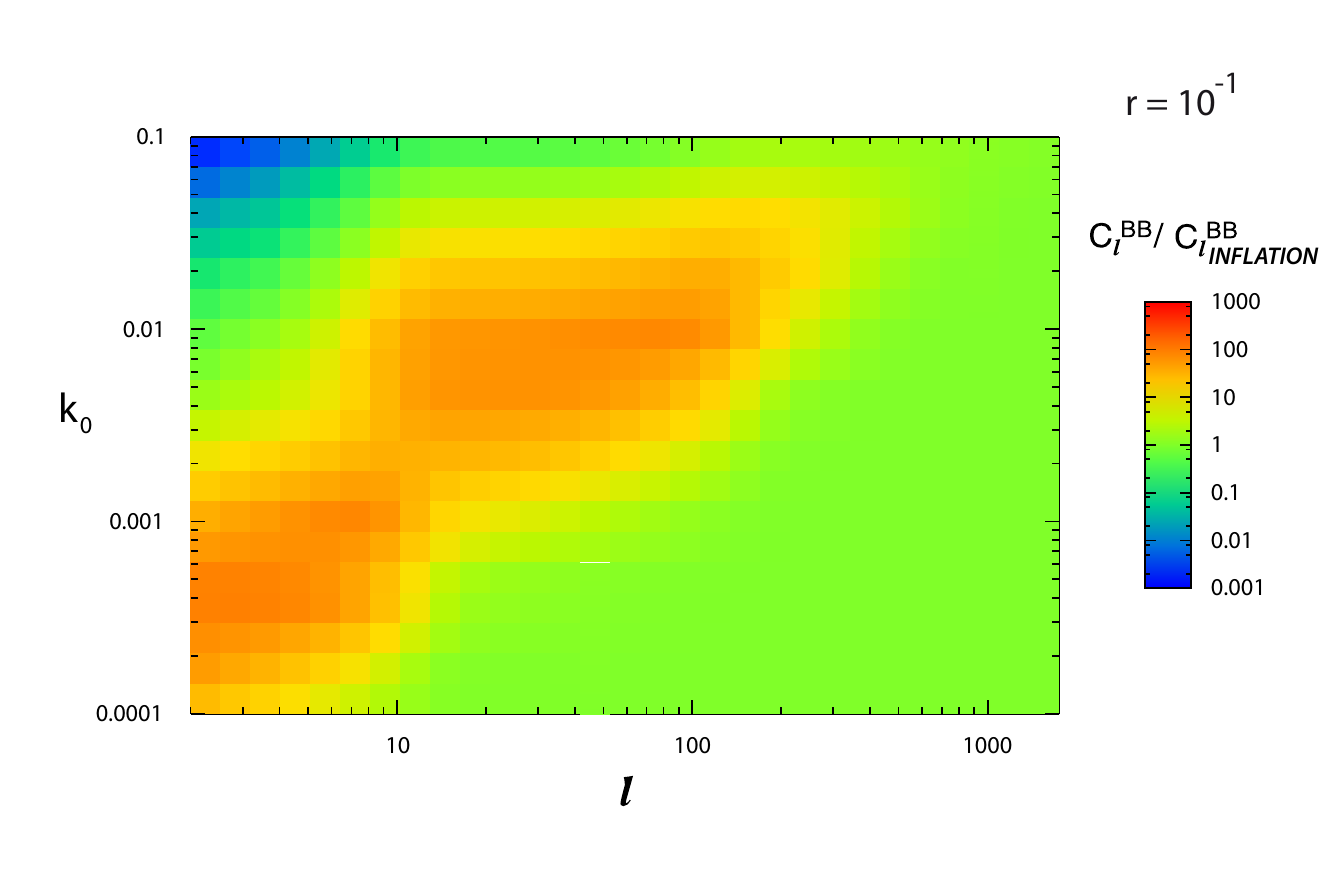}
\caption{Angular power spectrum of the B modes divided by the standard inflationary prediction as a function of $k_0$ with a 
tensor-scalar ratio of $10^{-1}$ and $R=500$.}
\label{BB-1i}
\end{figure}

\begin{figure}[htbp]
\centering
\includegraphics[width=0.5\textwidth]{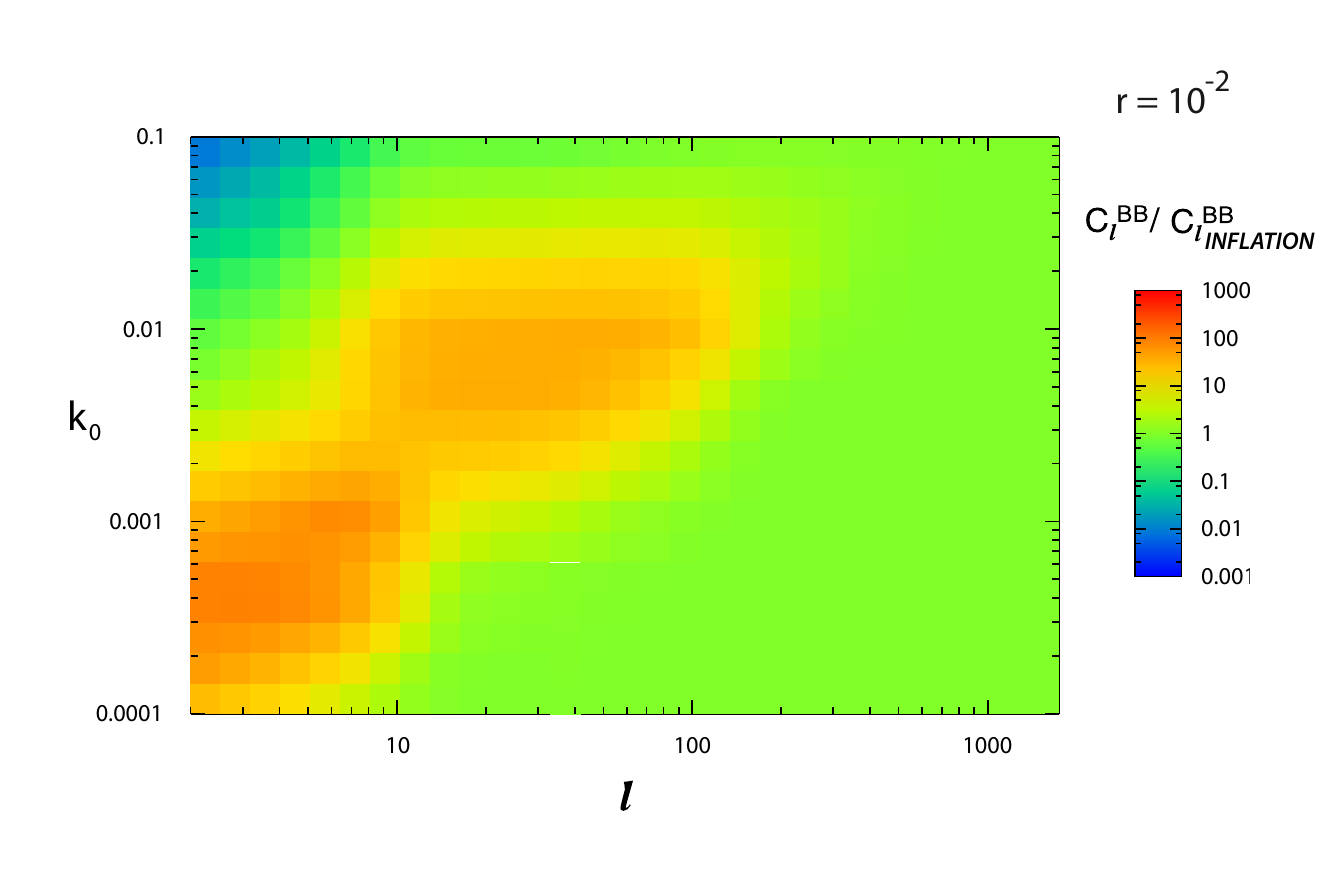}
\caption{Angular power spectrum of the B modes divided by the standard inflationary prediction as a function of $k_0$ 
with a tensor-scalar ratio of $10^{-2}$ and $R=500$.}
\label{BB-2i}
\end{figure}

\begin{figure}[htbp]
\centering
\includegraphics[width=0.4\textwidth]{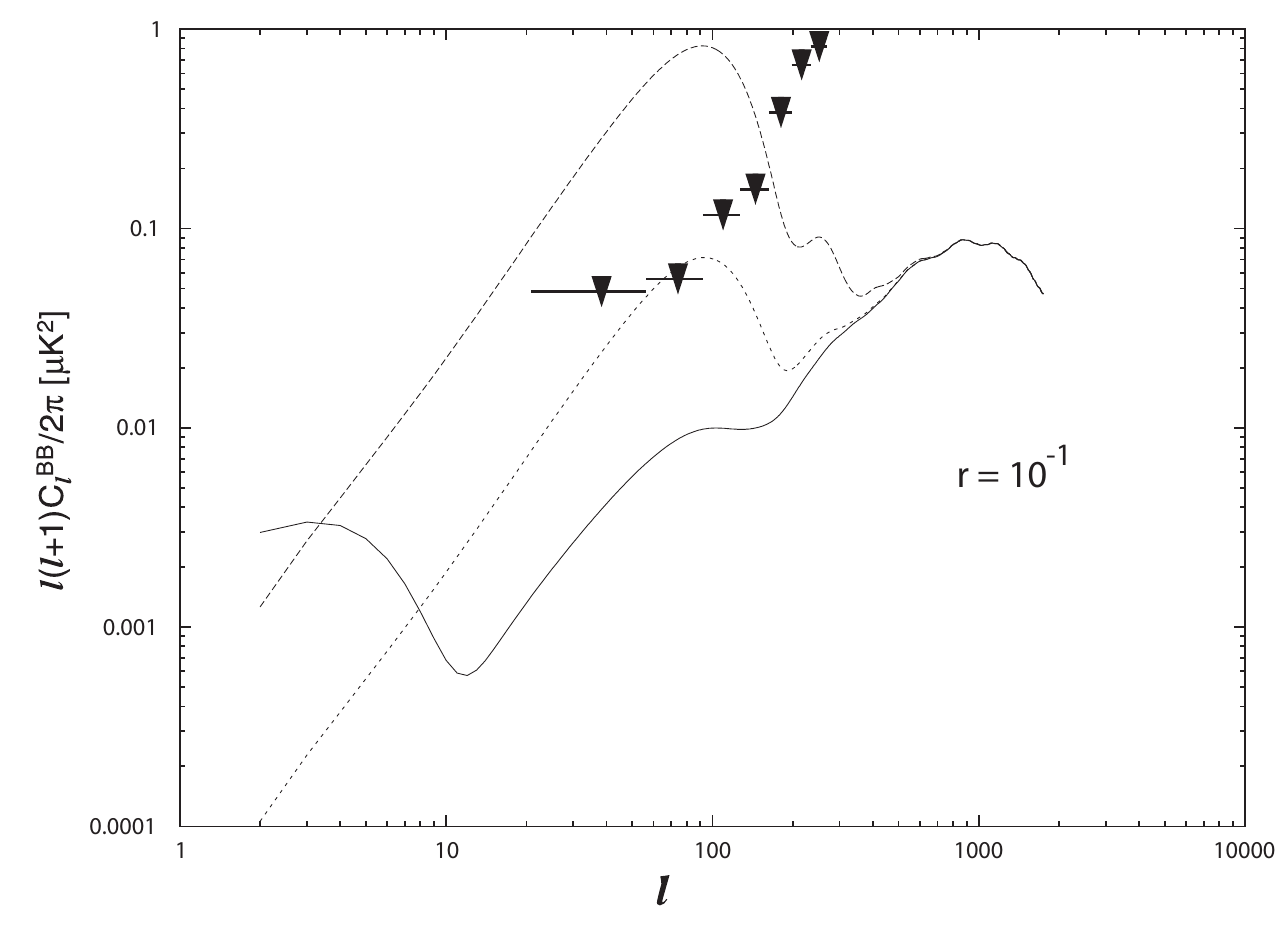}
\caption{B modes for $R=50$ (dashed), $R=500$ (long dashed) and $k_0 = 10^{-2}$ for a tensor-scalar ratio of $10^{-1}$. The solid line corresponds to the prediction of inflation with a tensor-scalar ratio of $10^{-1}$. The $95\%$ confidence upper limits on the detection of B-modes of polarization given by BICEP \cite{bicep} are also pictured.}
\label{BBn1}
\end{figure}

\begin{figure}[htbp]
\centering
\includegraphics[width=0.4\textwidth]{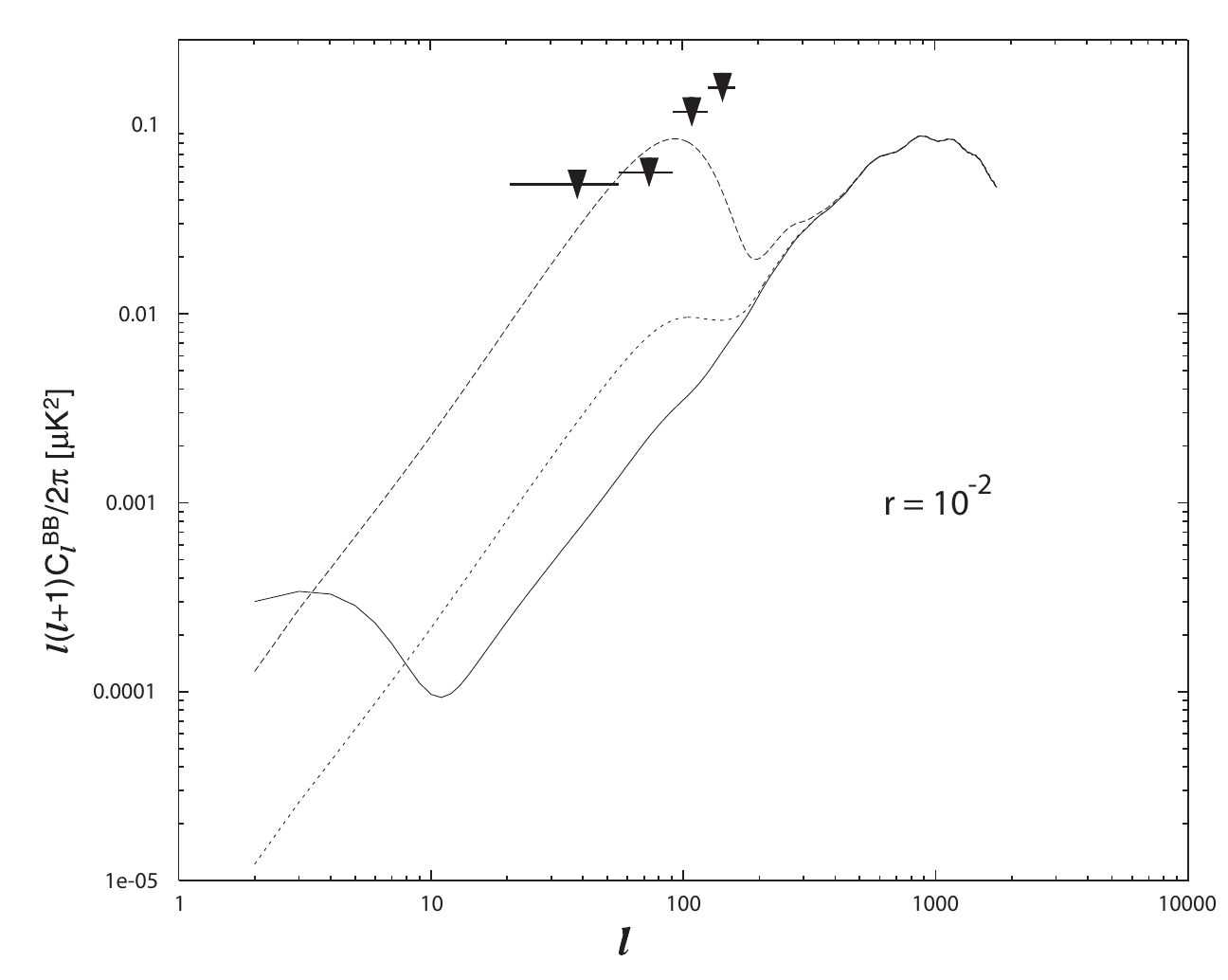}
\caption{B modes for $R=50$ (dashed), $R=500$ (long dashed) and $k_0 = 10^{-2}$ for a tensor-scalar ratio of $10^{-2}$. The solid line corresponds to the prediction of inflation with a tensor-scalar ratio of $10^{-2}$. The $95\%$ confidence upper limits on the detection of B-modes of polarization given by BICEP \cite{bicep} are also pictured.}
\label{BBn2}
\end{figure}

An important question is whether future experiments would be able to discriminate between B-modes produced by inflation and B-modes arising from a 
bouncing cosmology in the framework of LQC. 
A global analysis on all multipoles of a similar model was performed in \cite{GrainBarrauCailleteauMielczarek} using a Fisher analysis method 
to determine the detectability of the bump of the primordial spectrum of gravitational waves. 
Here we will take a complementary point of view, we will determine the multipoles where the B-modes arising from LQC can be distinguished 
from those produced by a standard inflationary scenario. 
In Fig. \ref{BB-1i} and \ref{BB-2i} are presented the B-mode power spectra divided by the standard inflationary prediction as a function 
of the parameter $k_0$ with tensor-scalar ratios of $10^{-1}$ and $10^{-2}$ respectively. 
Due to the contribution of the lensing, the LQC and inflationary spectra are identical at small angular scales. However, for multipoles 
smaller than $200$ the LQC spectra show significant departures from the standard inflationary prediction for identical tensor-scalar ratios 
in the two regions described previously. 
For $R=500$, which corresponds to the primordial spectrum of Fig. \ref{big9} with $F_B = 10^{-9}$, those departures can reach two 
orders of magnitude for $k_0 \approx 10^{-2}$. 
The corresponding angular power spectra are presented in Fig. \ref{BBn1} and \ref{BBn2}.
In the following we will only focus on spectra obtained for $k_0 = 10^{-2}$ since it corresponds to the region 
of the parameter space where the models can be most easily distinguished.

\begin{figure}
\centering
\includegraphics[width=0.4\textwidth]{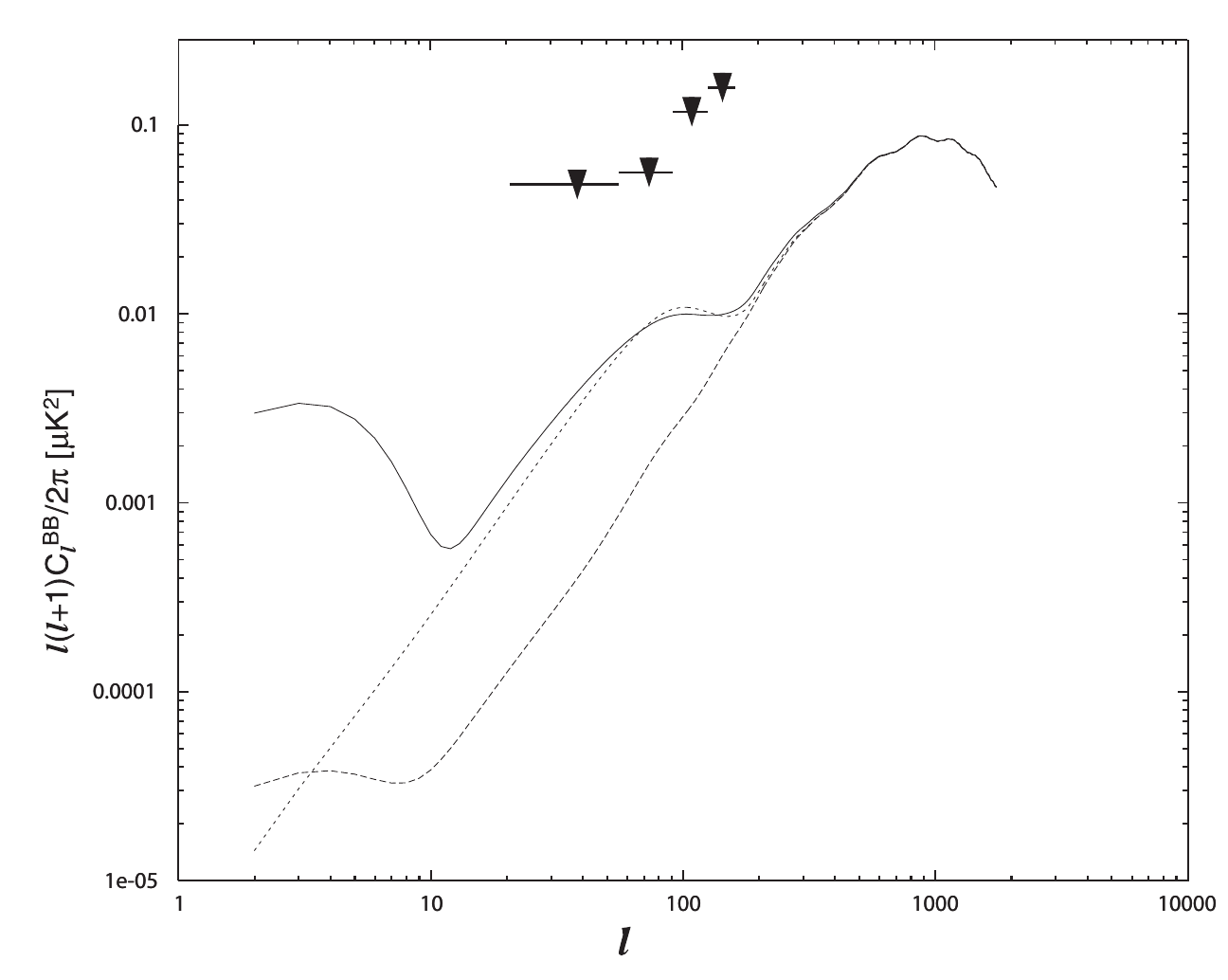}
\caption{B modes for $R=500$ (long dashed) and $k_0 = 10^{-2}$ for a tensor-scalar ratio of $10^{-3}$. The solid and dashed lines correspond respectively to the predictions of inflation with tensor-scalar ratios of $10^{-1}$ and $10^{-3}$. The $95\%$ confidence upper limits on the detection of B-modes of polarization given by BICEP \cite{bicep} are also pictured.}
\label{BBn3}
\end{figure}

For high values of $R$ the spectra are very different from the inflationary spectra as they possess a peak around $\ell = 100$ which should allow to discriminate between the models. 
In fact in the case of a tensor-scalar ratio $r$ of $10^{-1}$, a large region of phase space is even already excluded by the BICEP data \cite{bicep} (see Fig. \ref{BBn1}).
For smaller tensor-scalar ratios like $r = 10^{-2}$ only high values of $R$ are excluded by the data. 
For smaller values of $R$ and thus for higher values of $F_B$ the spectra come closer to the inflationary prediction in the observable range ($\ell > 20$), especially for low values of the tensor-scalar ratio.
Nevertheless, the importance of the backreaction \cite{backreaction2} grows as $F_B$ gets bigger and the effective model of LQC that we are using here is no more valid.
Therefore our conclusions only concern relatively large values of R.

The situation changes as the tensor-scalar ratio decreases and the inflationary and LQC spectra might be indistinguishable. 
In Fig. \ref{BBn3} not only is the gap between the inflationary and LQC spectra for $r=10^{-3}$ much smaller than in the $r=10^{-1}$ or $r=10^{-2}$ cases 
but the LQC spectrum for $r=10^{-3}$ also yields the same B-modes as inflation with $r=10^{-1}$ in the observable range ($\ell > 20$). 
This arises from the fact that the LQC spectra no longer possess a peak at low tensor-scalar ratios but a plateau around $\ell = 100$ which is 
similar to the one present in the inflationary spectrum.  
The B-modes of polarization produced in the frame work of LQC should thus be distinguishable in the observable range of multipoles from the 
inflationary prediction for low values of the potential energy at the bounce and for moderate tensor-scalar ratios. 
Therefore, future satellite and ground-based experiments should either be able to discriminate between LQC and inflation or to strongly 
constrain the effective model of LQC used in this study.


\end{document}